Forecasting the 2017-2018 Yemen Cholera Outbreak with Machine Learning

Rohil Badkundri[1*]

Victor Valbuena[1]

Srikusmanjali Pinnamareddy[1]

Brittney Cantrell[1]

Janet Standeven[1]

[1]Lambert High School

*805 Nichols Rd, Suwanee, GA 30024; rohilbadkundri@gmail.com; 404-398-5020

Abstract Word Count: 243

Text Word Count: 3431

Figures, Tables, and Supplementary Information Count: 1 table, 2 manuscript figures, 2 supplementary figures

## Abstract

The ongoing Yemeni cholera outbreak has been deemed one of the worst cholera outbreaks in history, with over a million people impacted and thousands dead. Triggered by a civil war, the outbreak has been shaped by various political, environmental, and epidemiological factors and continues to worsen. While cholera has several effective treatments, the untimely and inefficient distribution of existing medicines has been the primary cause of cholera mortality. With the hope of facilitating resource allocation, various mathematical models have been created to track the Yemeni outbreak and identify at-risk administrative divisions, called governorates. Existing models are not powerful enough to accurately and consistently forecast cholera cases per governorate over multiple timeframes. To address the need for a complex, reliable model, we offer the Cholera Artificial Learning Model (CALM); a system of four extreme-gradient-boosting (XGBoost) machine learning models that forecast the number of new cholera cases a Yemeni governorate will experience from a time range of 2 weeks to 2 months. CALM provides a novel machine learning approach that makes use of rainfall data, past cholera cases and deaths data, civil war fatalities, and inter-governorate interactions represented across multiple time frames. Additionally, the use of machine learning, along with extensive feature engineering, allows CALM to easily learn complex non-linear relations apparent in an epidemiological phenomenon. CALM is able to forecast cholera incidence 2 weeks to two months in advance within a margin of just 5 cholera cases per 10,000 people in real-world simulation.

# Introduction

## Cholera in Yemen

Cholera is a waterborne disease caused by the bacterium *Vibrio cholerae,* which has plagued mankind for centuries, and continues to do so despite the advances of modern medicine. The ongoing cholera outbreak in Yemen, which began in October of 2016, has been deemed "the largest documented cholera outbreak" through a comprehensive analysis of cholera surveillance data (Camacho et al., 2018). Triggered by a devastating civil war, cholera has spread rampantly across the country, with WHO weekly bulletins reporting 1,055,788 suspected, 612,703 confirmed cases of cholera, and 2,255 confirmed deaths as of April 2017 (World Health Organization, 2017). While cholera has several effective treatments, including Oral Cholera Vaccinations (OCVs) reported to work with an 80.2% prevention rate, the inefficient and untimely distribution of medicine has been the primary cause of cholera mortality (Azman et al., 2016; Camacho et al., 2018). The Yemeni outbreak has been largely sporadic, occuring in waves spawned by a variety of environmental, political, and epidemiological factors, including rainfall, civil war conflict, and cholera incidence and mortality (Camacho et al., 2018). Studies suggest a third wave of cholera transmission may resurge during the rainy season of 2019, resulting in an urgent need for a forecast that details precisely when, where, and how many people could potentially contract the disease (Camacho et al., 2018). With a comprehensive, actionable forecast, health organizations have the opportunity to deploy prevention methods in a highly targeted, efficient fashion, allowing for the mitigation of the outbreak (Camacho et al., 2018).

## Our Solution

We present CALM, the Cholera Artificial Learning Model, a system of four extreme-gradient-boosting (XGBoost) machine learning models that, working together, forecast the number of new cholera cases any given Yemeni governorate will experience for multiple time intervals ranging from 2 weeks to 2 months. Specifically, we provide models that predict new cholera cases in two week time frames 0-2 weeks from present, 2-4 weeks from present, 4-6 weeks from present, and 6-8 weeks from present. With extensive engineering of predictive features, the models utilize a large span of relevant datasets, including rainfall, past cholera incidence and mortality, and civil war mortalities. By predicting numbers of new cases per 10,000 people that each governorate will experience across multiple time frames, CALM provides a comprehensive and accurate forecast of the Yemen cholera outbreak, allowing for necessary preventative action to be taken. Furthermore, the geographic divisions (governorates) for which incidence are predicted are specific enough that practical measures can be taken to distribute medicines to those in need. For reference, YE-AM (Amran), the governorate with the greatest cumulative cholera case count normalized by population, has an area of 9,587 square kilometers (Yemen, 2014).

**Uniqueness of Approach**

Several cholera-predicting models have been constructed in the past, particularly in areas where cholera is seasonal and non-sporadic, such as Bangladesh. Because of seasonal patterns of cholera in these nations, these models tend to be simpler, often modeling linear relationships between environmental factors and cholera (Jutla, Akanda, Unnikrishnan, Huq, & Colwell, 2015; Jutla, Akanda, & Islam, 2010). In the case of Yemen, however the ongoing civil war presents a significant non-seasonal and sporadic influence, necessitating a more complex model capable of capturing this nonlinear influence. Our extreme gradient boosting approach provides this, offering a robust, principled approach used widely by data scientists to achieve state-of-the-art results on many machine learning challenges (Chen & Guestrin, 2016). Through not only

incorporating environmental and epidemiological factors, but also civil war data, we make use of the broad range of cholera-impacting factors in Yemen. Moreover, by employing robust machine learning strategies, our model is able to derive a deeper understanding of this data and ultimately strongly model the nonlinear trends of cholera in Yemen. It should be noted that Dr. Antarpreet Jutla has recently constructed a model for Yemen, but it is only available via press release, making direct comparison again difficult (Cole, 2018). Still Jutla's most recent model assesses general cholera risk, whereas CALM forecasts new cholera cases across multiple geographical and temporal scales, providing a more comprehensive report of coming outbreaks.

## Materials and Methods

### Cholera Epidemiological Data

Cholera case and death statistics are reported by the WHO, where health experts work directly with Yemeni health authorities at both the country and local levels to record all reported cholera cases and deaths (WHO presence in Yemen). The data, collected by the WHO, was accessed through the Humanitarian Data Exchange, which provided reports of accumulated new cholera cases and deaths per governorate from up to May 22, 2017, to February 18, 2018. Past cholera cases and deaths were included in the model with the simple assumption that they would be predictive of future cases, as cholera is spread from person to person through contaminated food or water sources.

### Rainfall Data

As *V. cholerae* is indigenous to aquatic environments, rainfall is a significant predictor of the transmission of cholera. In areas exposed to heavy rainfall, through the collapse of sanitary and health infrastructure, interaction between contaminated water and human activities accelerates, resulting in the potential for cholera infection (Jutla et al., 2013). In fact, an analysis of surveillance date for the Yemen outbreak from 2016 to 2018 found the relative risk of cholera 10 days after a weekly rainfall of 25 mm was 42% higher than compared with a week without rainfall (Camacho et al., 2018). At the same time, during the drought season, the use of unsafe water sources is hypothesized to contribute to the increasing levels of cholera as well (Camacho et al., 2018), and so rainfall became an important data source.

Daily rainfall data for Yemen from January 1, 2017 to March 30, 2018 was accessed through NASA's Goddard Earth Sciences Data and Information Services Center (GES DISC), which provides Global Precipitation Measurement (GPM) data. GPM data originates from an international network of satellites that use microwave imagers and precipitation radars to measure rainfall volume in various regions of the world (GPM, 2011). As there were individual data points for every .25 degrees of both latitude and longitude, reverse geolocation was performed to match coordinates with corresponding Yemeni governorates, and rainfall aggregates were taken over the multiple coordinates.

**Conflict Data for the Yemeni Civil War**

The collapse of Yemen's health, water, and sanitation sectors amidst the ongoing civil war has fueled the spread of cholera across the country (Camacho et al, 2018; Yemen's Cholera Crisis: Fighting Disease During Armed Conflict, 2017; Yemen: The Forgotten War, 2018). With attacks against hospitals and water supplies, the conflict has dissolved 55% of the country's medical, wastewater, and solid waste management infrastructure, making access to clean water and healthcare difficult and expensive (Camacho et al, 2018; Yemen's Cholera Crisis: Fighting Disease During Armed Conflict, 2017; Yemen: The Forgotten War, 2018). Information regarding the status of ongoing conflicts, namely the severity in terms of death toll, was collected to estimate a region's infrastructure ability to provide treatments in cholera in the following weeks. We retrieved our data from the Armed Conflict Location and Event Data Project (ACLED), which reported on the agents, locations, dates, and other characteristics of the politically charged conflict from January 1, 2016, to June 6, 2018 (Raleigh & Dowd, 2017). The number of daily casualties due to conflict in each Yemeni governorate was used as a metric for civil war related violence.

**Dataset Preparation**

With the objective of predicting new cholera cases in any given governorate in Yemen from week to week, there were a number of steps taken to prepare our data. To build a model more sensitive to sharp outbreaks and changes in cholera patterns, the case and death report time series were converted from total case/death reports to new case/death reports. This first involved calculating the number of new cholera cases/deaths that had occured between each WHO report. Next, as time between reports varied anywhere from a day to near a week, we linearly-interpolated our new cholera case and death calculations to arrive at daily new case and death data (e.g if 10 new cases occurred over a period of two days, we assumed 5 new cases occurred on each day). It should be noted that the country of Yemen encompasses 21 governorates or administrative divisions. While the CALM models were trained on data from all 21 governorates, data preparation on each governorate was performed separately to preserve each governorate's unique time series. Likewise, all values were normalized by the population of each governorate (e.g new cases per 10,000 people). Finally, we calculated our four target variables: the number of new cholera cases 0-2 weeks from the present day, 2-4 weeks from the present, 4-6 weeks from the present, and 6-8 weeks from the present.

*Figure 1: Diagram of Rolling-Window Cross-Validation*

Rolling-window cross-validation operating on a sliding scale. A rolling-window method was used to preserve the inherent temporal order of the data and prevent lookahead bias - the use of data that would not have been available during the period being analyzed to make predictions - which can create inaccurate results

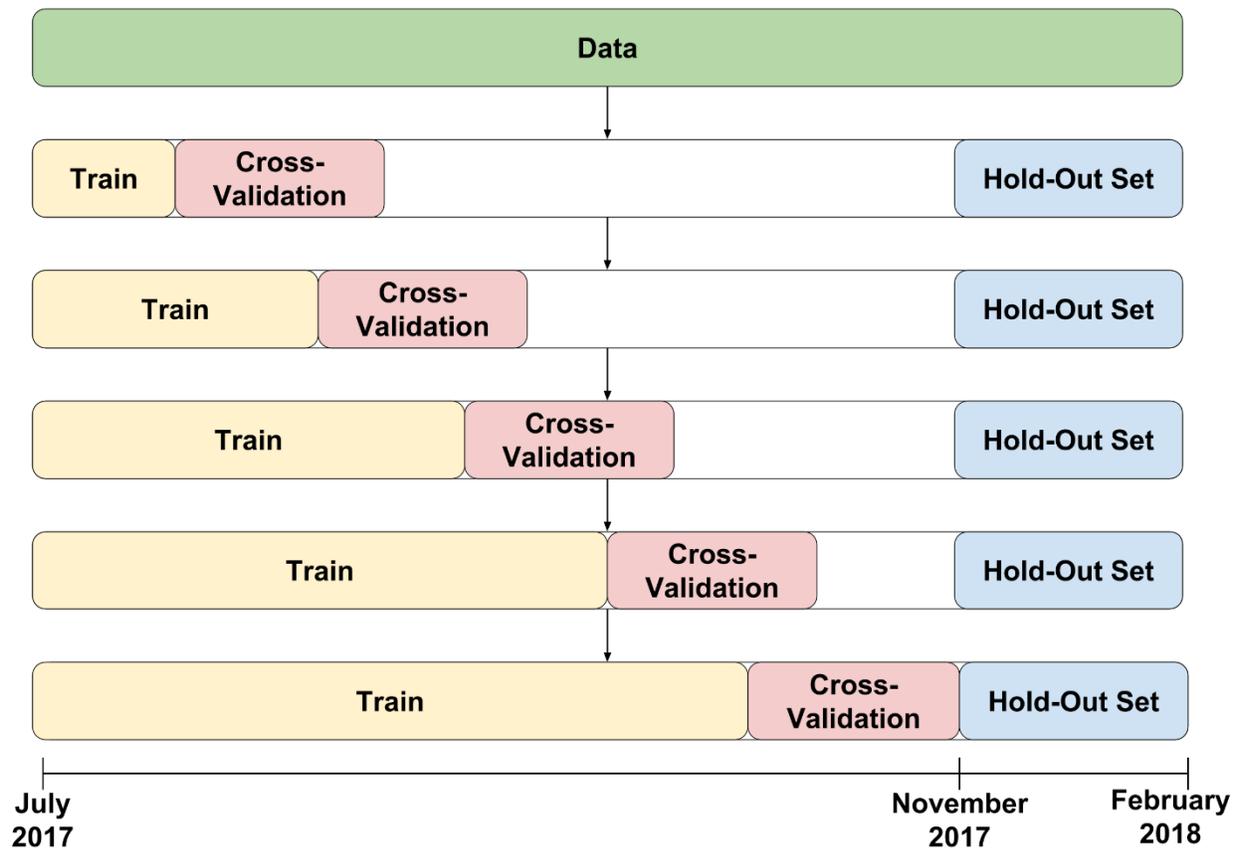

Our dataset was split into three portions: training, cross-validation, and a hold-out test set. The hold-out set was left untouched until the completion of our methods to provide a real-world simulation of our models' performance. For cross-validation, we utilizing a rolling-window method to engineer features and find optimal hyperparameters. Standard k-fold cross-validation involves randomly splitting the cross-validation set into N segments, then training N models such that the ith model uses the ith data segment as a set-aside validation set. This traditional method of cross-validation relies on the assumption that there is no relationship between observations in the data, meaning each observation is independent. However, with time-series data, there is an inherent temporal relationship between observations. Therefore, the data cannot be split up into random groups - the temporal order must be respected. In fact, using traditional k-fold cross-validation can result in dangerous look-ahead bias, which involves using data that would not have been available during the period being analyzed to make predictions, creating inaccurate results. Rolling-window cross-validation addresses this problem by preserving the temporal order of the dataset: with each cross-validation fold the training set is iteratively increased and the cross-validation set is moved to a future date (see Figure 1). Our base training set was defined from July 1, 2017 to August 15, 2017. While WHO reports extended back as far as May 22, 2017, we chose to start on July 1 in order to have enough prior data for feature calculation. Our cross-validation dataset was defined from August 15 to November 10, 2017. Finally, our hold-out set started from November 11 and extended to a final date in January/February, 2018, which varied for each defined target variable depending on the respective range (a 6-8 week forecast implies a larger time frame between current and forecast date than a 2-4 week forecast, and so the 6-8 week forecast holdout set would end prior to the 2-4 week forecast). Our five cross-validation sets were from August 16-31, August 31 to September 15, September 15-30, September 30 to October 15, and October 15-30. The final fold included data from October 30 to November 10, 2018 as a prediction set, but this does not cross into the hold-out set.

**Feature Engineering**

Given our lack of an expansive dataset, instead of an end-to-end feature learning process, we underwent an exhaustive feature extraction and selection process in order to arrive at our final features. Feature were derived from variable-length time-frames: 8 weeks, 6 weeks, 4 weeks, 2 weeks, and 1 week prior from the data on which future predictions would be made. The process of feature derivation was the same for each time-length. To describe phenomenon in geographically neighboring governorates, we took the mean cholera cases, rainfall, etc. in neighboring areas and included this value as a separate timeseries. We extracted 45,000 potentially relevant features using the tsFresh (Time Series Feature extraction based on scalable hypothesis tests) package. Given some dataset, tsFresh automatically calculates hundreds of unique time-series features. These features range from those commonly used by data scientists on time-series data (mean, max, number of peaks, etc.) to more complicated features (Christ et al., 2018). Through covering this expansive set of features, we increased the odds of capturing ideal representations of our data. To remove features containing redundancies, noise, or irrelevant information, we first utilized the tsFresh package's scalable hypothesis test function, which statistically evaluates the importance and explaining-power of each feature using hypothesis tests. For real-valued features, the Kendall rank correlation coefficient was calculated, and for binary features, the Mann–Whitney U test was used. After a multiple test procedure, the hypothesis tests provide a q-value indicating the statistical significance of the feature. With a q-value cutoff of 0.001, we found four sets of statistically significant features ~15,000 in number for each time-frame prediction. Next, we removed features that were greater than 97% correlated with each other, as these features would be redundant to our model, thus providing us with sets of ~10,000 features to further narrow. We trained and tuned an XGBoost model to rank the features in order of importance for each time-range prediction. Utilizing the ranking produced, we recursively added features based on if they added to our cross-validation loss, the root mean square error across all five cross-validation folds. This allowed us to arrive at the best 30-50 features for each time-range model.

**Model**

We utilized XGBoost, a random forest-based, extreme gradient boosting algorithm, to construct each of our models. Through bootstrap aggregation, the construction of often hundreds of decision trees that are trained on random subsets of the data and then collectively vote for the final prediction, XGBoost is able to address variance-related error (overfitting). XGBoost also addresses the converse, bias-related error (underfitting), through gradient boosting: the process by which each decision tree is constructed with a greater focus on the samples the prior trees had difficulties with (Chen & Guestrin, 2016). As opposed to simpler regression techniques utilized by previous models, XGBoost is able to gain a far deeper understanding of the data through nonlinear relations, while being able to distinguish from noise, making it an ultimately more robust choice of algorithm.

**Tuning**

We utilized Bayesian optimization to find optimal hyperparameters for our model. Hyperparameters were found for each time-range model individually. Specifically, we used the hyperopt package, which makes use of the tree parzen estimator to perform Bayesian optimization. We chose to minimize the average mean squared error across 5 cross-validation folds as our objective function. In contrast with a brute-force search over a defined set of hyperparameters, Bayesian optimization tracks prior evaluations to form probabilistic assumptions on an objective function given a set of hyperparameters, allowing informed choices to be made on which hyperparameters to try (Snoek et al., 2012). This allowed us to converge at optimal hyperparameters with far greater efficiency.

## Results

*Table 1: Cross-validation and Holdout Error for four XGBoost forecasting models*

Cross-validation and hold-out error for each of our four models. We defined error as the root mean squared error, measured in new cholera cases per 10,000 people. CV-error was obtained by taking the root of the mean of the model's performance across five rolling-window cross validation folds. Holdout-error was obtained by calculating the root mean squared error for predictions only on the holdout set.

| Forecast Range | XGBoost CV-error | XGBoost Holdout-error | Linear Regression Holdout-error | Linear Regression CV-error |
|---|---|---|---|---|
| *0 - 2 weeks* | 6.867 | 4.333 | 7.825 | 4.603 |
| *2 - 4 weeks* | 5.574 | 4.390 | 9.258 | 6.716 |
| *4 - 6 weeks* | 4.134 | 4.654 | 8.752 | 10.538 |
| *6 - 8 weeks* | 3.861 | 4.660 | 8.730 | 6.935 |

Our models are able to predict the number of new cholera cases any given governorate in Yemen will experience across multiple two-week intervals, with all of our models being able to predict within a margin of 5 cholera cases per 10,000 people in the hold-out set (see Table 1). In addition, our XGBoost models outperform baseline linear regression models across all benchmarks tested. Hold-out error represents our model's performance in real-world simulation, as the hold-out dataset was left untouched until final model evaluation. Our cross-validation error, similarly low, represents our model's performance on a reliable, but not entirely untouched dataset, as the cross-validation dataset was used for hyperparameter tuning and feature selection.

Keywords: cholera, yemen, machine learning, time series forecasting, xgboost

The mean number of cases any given governorate in Yemen experienced within a two-week span was approximately 19.148, with the standard deviation being 21.311. As, in real-world simulation, all four of our predictive models are able to predict around ⅓ of a standard deviation of the number cases, our predictions are robust and reliable across all time frames. However, as our predictive timeframe passes farther into the future, the cross-validation error decreases and the hold-out error increases (see Table 1). This could be seen as a sign of marginal overfitting, but can also be attributed to the time-shift in data as the predictive range increases: at any given time, the 6-8 week forecast model is making predictions on a time-frame the 0-2 week model will predict on four weeks later. Cholera 6-8 weeks ahead of a given date can look different than 1-2 weeks ahead, and this change in data could be the cause of this unexpected decrease in cross-validation error.

*Figure 2: XGBoost Predictions of New Cases 0 to 2, 2 to 4, 4 to 6, and 6 to 8 Weeks in Advance for Five Governorates*

Forecasts for each time frame vs a sliding window of real cases. Data point represents new cholera cases in a two-week interval corresponding to the forecast range. True values: red; cross-validation predictions: green; hold-out predictions: blue.

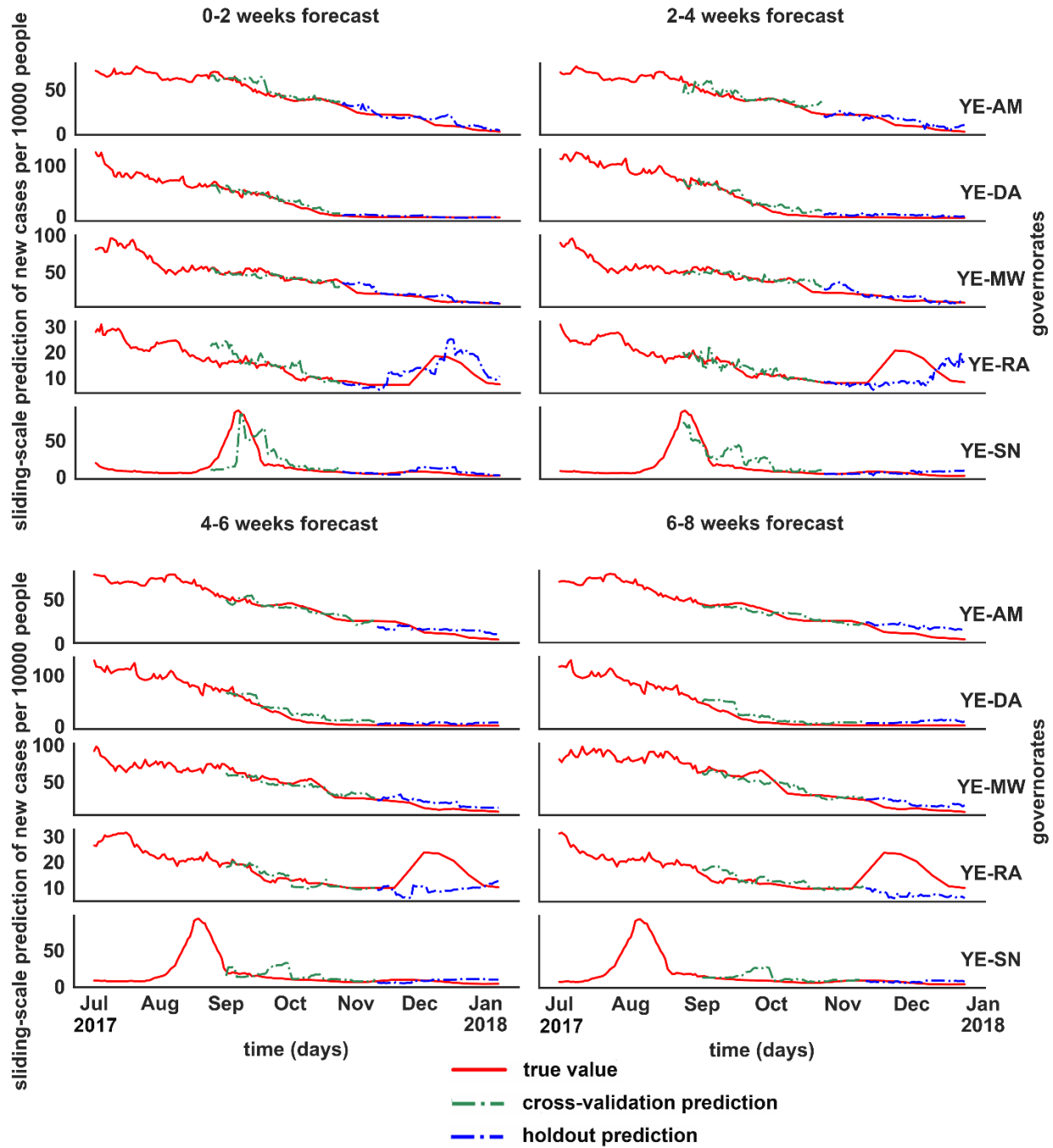

Our four-model system is able to accurately and comprehensively forecast cholera outbreaks across 21 governorates exhibiting heterogeneous behaviors. In Figure 2, five governorates were chosen to represent the entire range of behavior exhibited by all 21 of Yemen's governorates. Figures for all 21 governorates (see Appendix A) and an animation (see Appendix B) are also provided. Amran (YE-AM), Dhale (YE-DA), and Al Mahwit (YE-MW) experienced the greatest cumulative number of cholera cases from May 22, 2017 to February 18, 2018, respectively, making them three of the four governorates most affected by cholera overall in the given timeframe. These three governorates exhibit behavior highly similar to the other governorates, albeit at a higher scale. Our accurate delineation of the outbreak across all four time frames in these three governorates (see Figure 2) shows the usefulness of our models in the general case. YE-SA and YE-RA (shown in the bottom two subplots in Figure 2) present a rare and interesting case, as both experienced a sudden outbreak. As our predictive range increases, it becomes more difficult to predict sudden spikes, due to either a lack of information many weeks prior or the events preceding a sharp outbreak not having occurred yet. As a result, longer range models seem to predict sharp outbreaks with a certain lag. This can be seen in Figure 2 with the 4-6 week model's forecasting of the YE-RA outbreak, which shows an upward trend, but not a full spike being predicted. However, this is where the combination of all four of our models becomes most useful. While long-range models cannot easily predict outbreaks, our shorter-range models are able to pick up the slack once the outbreak becomes closer. Specifically, as seen in Figure 2, our 0-2 week forecasting model is able to predict incidence spikes in YE-RA and YE-SA, so even if our long-range models were unable to predict the outbreak immediately, we would still detect the outbreak at a later date.

## Discussion

### Summary

Cholera has killed millions of people, and without proper action, could continue to do so for years. Many have modeled and analyzed cholera outbreaks to forecast cholera risk or cases during outbreaks. Our system, CALM, the Cholera Artificial Learning Model, forecasts new cholera incidence, with the proof-of-concept forecasting cases in Yemen. Consisting of four XGBoost models working in conjunction, CALM is able to make predictions that are highly accurate and reliable, utilizing a broad range of predictive features, and robust machine learning techniques. In addition, by using a system of four models, CALM is not only able to provide a comprehensive 2 month report with 2-week intervals, but also ensures that the mistakes of longe-range model can be corrected by short-range ones. Sudden spikes in the cholera prediction may not be modeled accurately by models of later time frames, but more immediate aspects of the model, such as the predictor for 0-2 weeks, can effectively account for sudden changes in the cholera cases. CALM has the potential to be a reliable, powerful way to reduce the number of people suffering from cholera by providing advance notice of outbreaks and allowing for the distribution of medical supplies to pre-empt an outbreak.

### CALMWatch, an SMS Bot

We have also developed an SMS bot utilizing the Twilio API and the Flask web framework for gathering health and sanitation data from areas affected by a cholera outbreak. Coined CALMWatch, this bot allows for a healthcare organization or government agency to distribute an SMS survey to a given population so that affected people can report data related to an ongoing cholera outbreak such as cleanliness of water sources, water storage, and waste management. This data can then be fed into the CALM model in real time to increase the accuracy of the model and increase the size of its databases.

**Further Development**

While the efficacy of the model has only been demonstrated in Yemen, it is expected that with further development and adaptation CALM will be used to predict cholera outbreaks around the world. As development on the project progresses, we hope to construct a fully autonomous web-based software platform comprising of data collection bots that collect data from major health and sanitation sources and CALMWatch surveys, and an online platform that coordinates global usage of the model so that users can share and distribute results and model improvements more easily. Finally, through the incorporation of additional datasets, such as algal blooms, migration data, and OCV campaign data, we hope to engineer more features for our models.

Acknowledgements

We thank Dr. Andre Esteva, Dr. Eric Mintz, and Chris Waites for their advisement on the construction of our model.

We thank the developers of RatWatch, an open-source SMS-based rat reporting service for the Atlanta area whose developers graciously allowed us to modify it for the purposes of our own SMS bot.

Furthermore, we thank Scott Henderson, Shane Matthews, Alex Morse, Jackson Morgan, Luigi Ray-Montanez, Winnie Luo, Michael Koohang, Kevin Alvarez, and TJ Muehleman for their guidance and recommendations through the Day One Georgia work sessions.

Finally, we thank Tanishk Sinha, Sahil Jain, and Shanthi Hegde for their work in collecting preliminary datasets.

Appendix A

*XGBoost Predictions of New Cases 0 to 2, 2 to 4, 4 to 6, and 6 to 8 Weeks in Advance for 21 Governorates*

Forecasts for each time frame vs a sliding window of real cases. Data point represents new cholera cases in a two-week interval corresponding to the forecast range. True values: red; cross-validation predictions: green; hold-out predictions: blue.

Keywords: cholera, yemen, machine learning, time series forecasting, xgboost

## 0-2 week forecast model

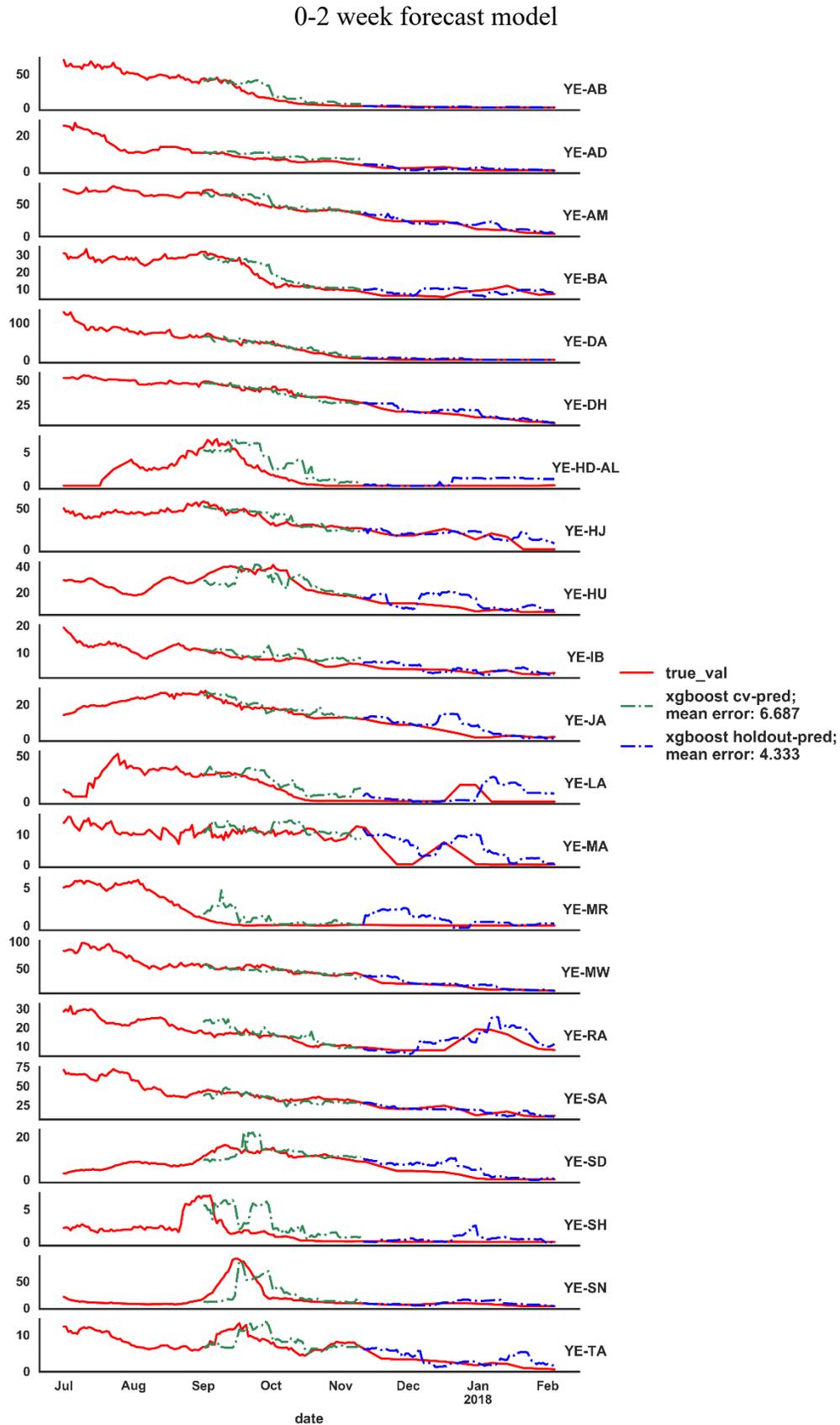

Keywords: cholera, yemen, machine learning, time series forecasting, xgboost

## 2-4 week model

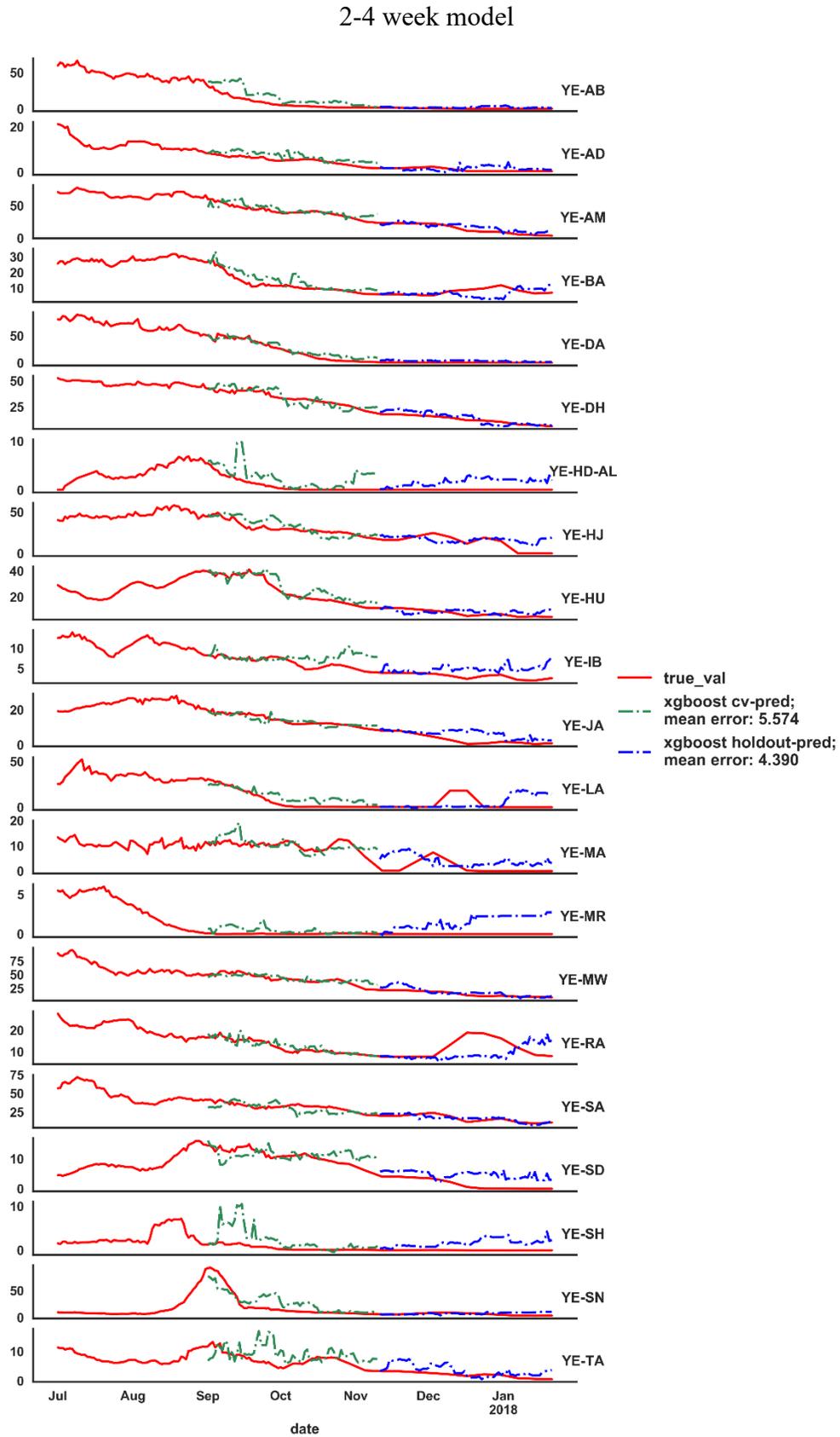

Keywords: cholera, yemen, machine learning, time series forecasting, xgboost

4-6 week model

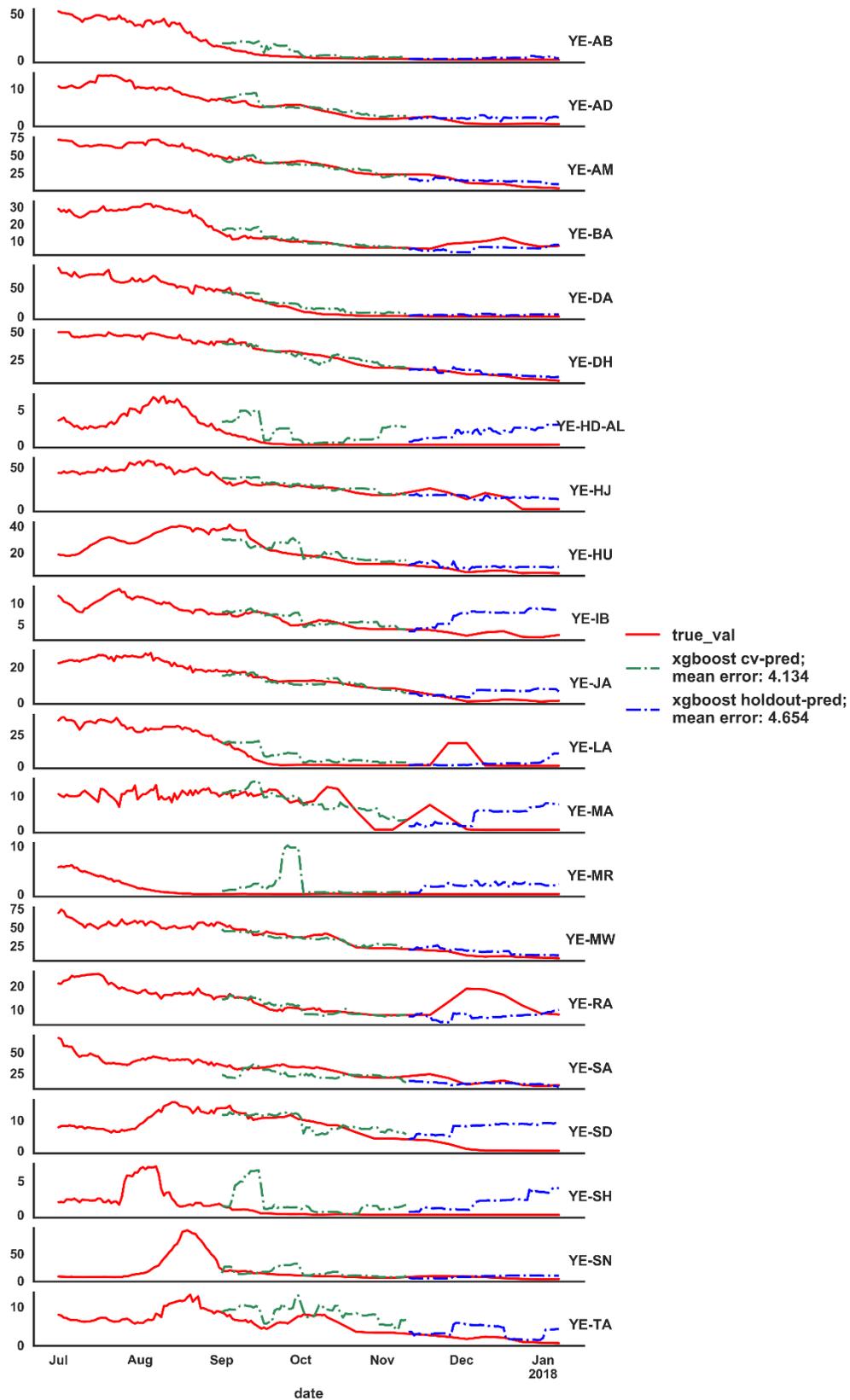

Keywords: cholera, yemen, machine learning, time series forecasting, xgboost

## 6-8 week model

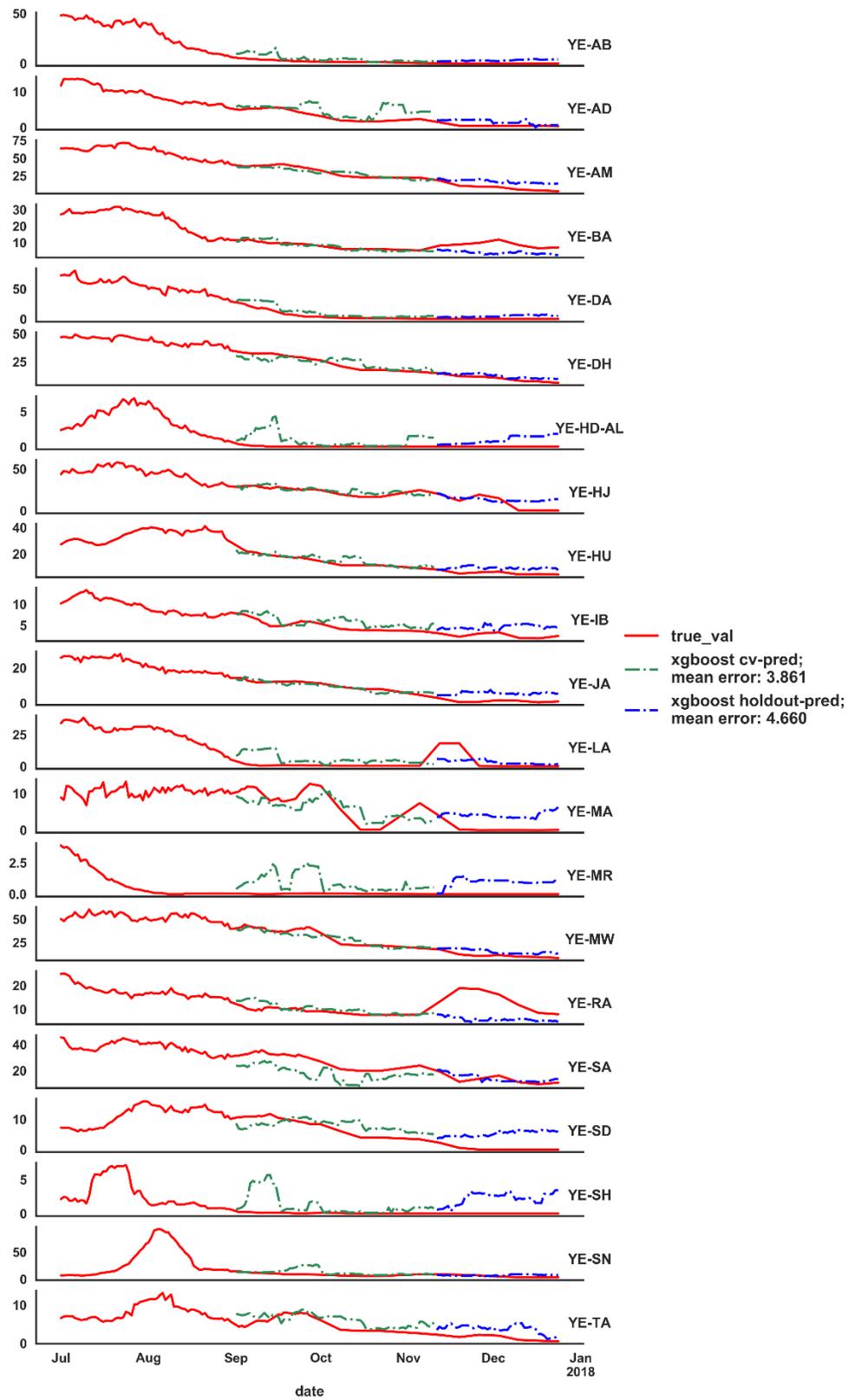

Appendix B

*Cumulative Cholera Cases for 5 Governorates.*

Animation visualizing our predictions against cumulative cholera case counts across time. Reverse differencing was performed to convert our predictions from new cases to total cases. X axis is time (days) and y axis is cumulative cholera cases per 10,000 people

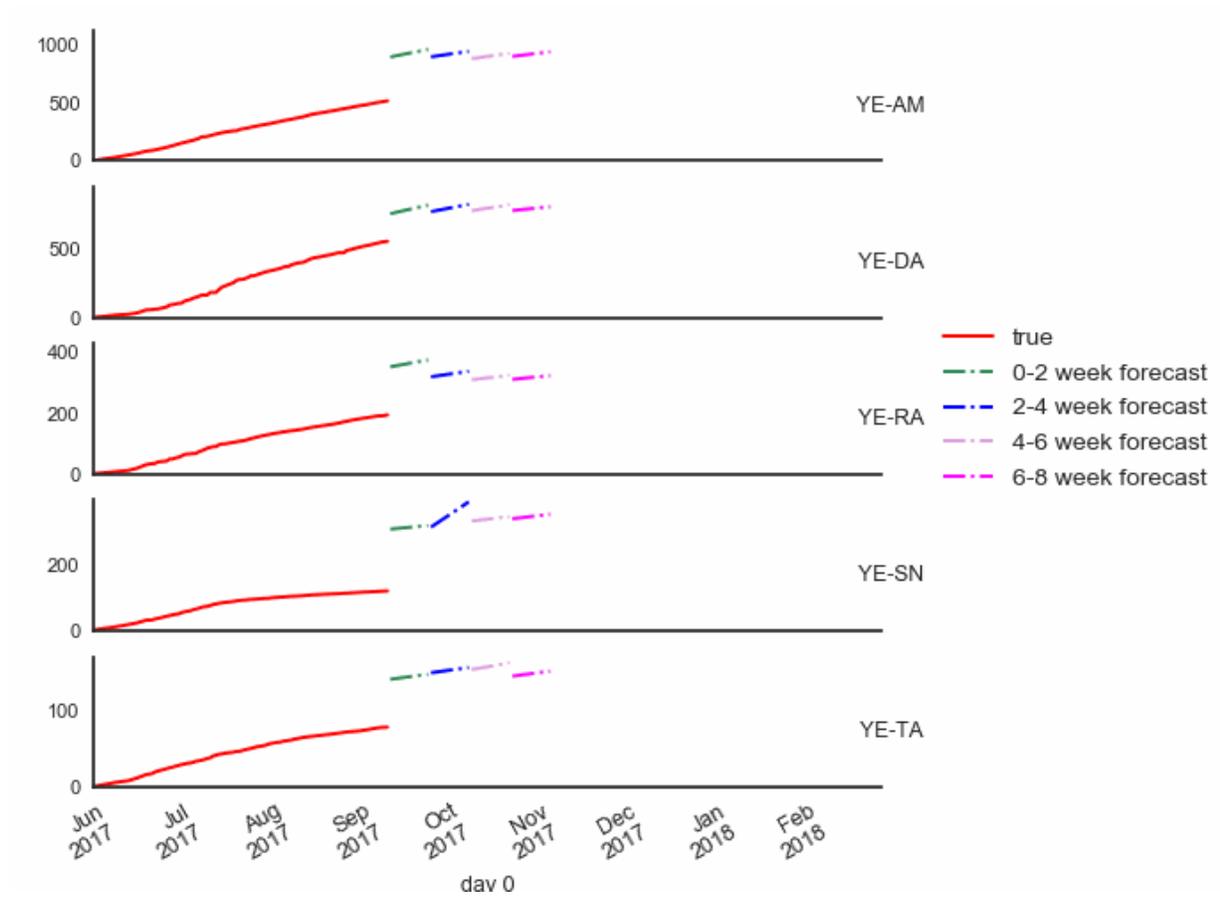

(Can be viewed online at http://2018.igem.org/Team:Lambert_GA/Software?#target10)